\documentclass[aps,prc,twocolumn,superscriptaddress,showpacs,floatfix,nofootinbib]{revtex4-1}
\usepackage{amsmath,graphicx}
\begin{document}

\title{Constraining the equation of state with identified 
  particle spectra}

\author{Akihiko Monnai}
\affiliation{
Institut de physique th\'eorique, Universit\'e Paris Saclay, CNRS, CEA, F-91191 Gif-sur-Yvette, France} 
\author{Jean-Yves Ollitrault}
\affiliation{
Institut de physique th\'eorique, Universit\'e Paris Saclay, CNRS, CEA, F-91191 Gif-sur-Yvette, France} 
\date{\today}

\begin{abstract}
We show that in a central nucleus-nucleus collision, the variation of
the mean transverse mass with the multiplicity is 
determined, up to a 
rescaling, by the variation of the energy over entropy ratio as a
function of the entropy density, thus providing a direct link between
experimental data and the equation of state. 
Each colliding energy thus probes the equation of state at an
effective entropy 
density, whose approximate value is $19$~fm$^{-3}$ for Au+Au collisions
at 200~GeV and $41$~fm$^{-3}$ for Pb+Pb collisions at 
2.76~TeV, corresponding to temperatures of $227$~MeV and $279$~MeV if
the equation of state is taken from lattice calculations.  
The relative change of the mean transverse mass as a function of the 
colliding energy gives a direct measure of the pressure over energy
density ratio $P/\epsilon$, at the corresponding effective density. 
Using RHIC and LHC data, we obtain $P/\epsilon=0.21\pm 0.10$, in
agreement with the lattice value $P/\epsilon=0.23$ in the corresponding
temperature range. 
Measurements over a wide range of colliding energies using a single detector
with good particle identification would help reducing the error. 
\end{abstract}

\maketitle

\section{Introduction}

One of the motivations for studying 
nucleus-nucleus collisions at high energies is to probe experimentally
the equation of state of QCD matter~\cite{Stoecker:1986ci}. 
Ultrarelativistic collisions probe the phase diagram at vanishing
chemical potential: at high temperatures, hadrons merge into a
quark-gluon plasma.   
It was originally hoped that this change occurred through a
first-order phase transition~\cite{Yaffe:1982qf}.  
However, it was progressively understood that it is a smooth, analytic 
crossover~\cite{Brown:1990ev,Aoki:2006we}, and that a phase
transition, if any~\cite{deForcrand:2006pv}, can only take place at high
baryon density~\cite{Fodor:2004nz,Fukushima:2010bq}. 
The equation of state of baryonless QCD matter is now known 
precisely from lattice simulations with 
physical quark masses~\cite{Borsanyi:2010cj,Bazavov:2014pvz}. 
The goal of this paper is to understand the imprints of the equation
of state on heavy-ion data, in particular transverse momentum
spectra. 

Relativistic hydrodynamics~\cite{Gale:2013da} plays a central role in
our understanding of heavy-ion observables in the soft sector.  
Its simplest version is ideal hydrodynamics~\cite{Kolb:2003dz}, which
describes most of the qualitative features seen in transverse
momentum spectra, elliptic flow, and interferometry
radii~\cite{Huovinen:2006jp}. This simple description can be refined
by taking into account finite-size corrections due to
viscosity~\cite{Romatschke:2009im} which are important for azimuthal
anisotropies~\cite{Heinz:2013th}. 
The equation of state lies at the core of the hydrodynamic
description, and the vast majority of modern hydrodynamic
calculations~\cite{Luzum:2008cw,Schenke:2010rr,Song:2010mg,
Bozek:2011ua,Hirano:2010jg,Petersen:2010cw,Karpenko:2010te,
Akamatsu:2013wyk,Holopainen:2010gz,Pang:2012he,Roy:2010qg},
which give a satisfactory description of soft observables, 
use as an input an equation of state from lattice QCD calculations. 

While the success of hydrodynamics suggests that 
equilibration takes place to some 
degree~\cite{Gelis:2013rba,Kurkela:2015qoa}, 
most dynamical calculations predict that the system produced in the
early stages of a heavy ion collision is far from chemical
equilibrium, typically with overpopulation in gluon
numbers~\cite{Blaizot:2011xf} and underpopulation in quark
numbers~\cite{Blaizot:2014jna,Monnai:2014xya}.  
The resulting effective equation of state might differ significantly
from that calculated in lattice QCD, and it is important to understand
what experimental data tell us about the equation of state,
beyond a comparison between different lattice
results~\cite{Dudek:2014qoa,Moreland:2015dvc}.
It has been recently shown that a simultaneous fit of several observables to
hydrodynamic calculations constrains the equation of state
to some extent~\cite{Pratt:2015zsa}.
However, this recent study uses a
systematic, Bayesian framework, and the nature of the 
relationships between model parameters and observables remains obscure. 
Further Bayesian studies have
shown~\cite{Sangaline:2015isa} that  
interferometry radii and transverse momentum spectra are the 
observables which are most sensitive to the equation of state, but
they are still unable to provide a simple picture of how this
dependence takes place. Another related approach is 
to use a deep learning method to distinguish the crossover and first-order
phase transitions in equations of state from heavy-ion particle 
spectra \cite{Pang:2016vdc}.

We show that for central collisions, the variation of the mean
transverse mass per particle as a function of the multiplicity density
$dN/dy$ (which itself depends on the collision energy $\sqrt{s}$) 
reproduces, up to proportionality factors, the variation of energy
over entropy ratio $\epsilon/s$ as a function of the entropy density
$s$~\cite{Blaizot:1987cc}.
We illustrate our point by discussing an ideal experiment in
Sec.~\ref{s:gedanken}. 
We then carry out detailed hydrodynamic simulations using a variety of
equations of state. 
The equations of state are presented in Sec.~\ref{s:eos}. 
Results from hydrodynamic calculations are discussed in
Sec.~\ref{s:meanmt}. 
Calculations are compared with experimental data from RHIC and LHC 
in Sec.~\ref{s:data}.

\section{An ideal experiment} 
\label{s:gedanken}

In order to illustrate our picture, we first describe a simple 
ideal experiment: the fluid is initially at rest in thermal
equilibrium at temperature $T_0$ in a container of arbitrary shape, and
large volume 
$V$. At $t=0$, the walls of the container disappear and the fluid
expands freely into the vacuum. If $V$ is large enough, 
this expansion follows the laws of ideal hydrodynamics. 
At some point, the fluid transforms into $N$ particles. We assume for
simplicity that this transformation occurs at a single freeze-out
temperature $T_f$~\cite{Cooper:1974mv}.

The thermodynamic properties at the initial
temperature $T_0$ can be easily be reconstructed by measuring the
energy $E$ and the number of particles $N$ at the end of the
evolution, provided that the initial volume $V$ is known. 
The total energy $E$ is conserved throughout the evolution,
hence the initial energy density is: 
\begin{equation}
\label{efinal}
\epsilon(T_0)=\frac{E}{V}.
\end{equation}
For simplicity, we assume throughout this paper that the net baryon
number is negligible (which corresponds to high-energy
collisions) so that the energy density depends solely on the
temperature. 

The initial entropy density can be inferred from the final number of
particles $N$. 
Ideal hydrodynamics conserves the total entropy $S$. 
The fluid is transformed into particles at the freeze-out temperature
$T_f$, and the multiplicity $N$ is directly proportional to the
entropy.\footnote{Both the multiplicity $N$ and the entropy $S$ are scalar quantities,
hence, the entropy per particle only depends on the freeze-out
temperature $T_f$, not on the fluid velocity.} 
Therefore, the initial
entropy density is related to the final multiplicity through the
relation:
\begin{equation}
\label{sfinal}
s(T_0)= \left(\frac{S}{N}\right)_{T_f}\frac{N}{V},
\end{equation}
The volume dependence cancels in the energy per particle: 
\begin{equation}
\label{decomposition}
\frac{\epsilon(T_0)}{s(T_0)}=\left(\frac{N}{S}\right)_{T_f}\frac{E}{N},
\end{equation}
One can repeat the experiment for several values of the initial 
density, and plot the energy per particle $E/N$ as a function of $N/V$. 
One thus obtains a plot of $\epsilon/s$ versus $s$, 
which gives access to the equation of state. 
Note that Eqs.~(\ref{sfinal}) and (\ref{decomposition}) do not involve
the fluid velocity pattern, 
which depends on the shape of the initial volume. 
Hydrodynamic modeling only enters through the entropy per particle at
freeze-out $(S/N)_{T_f}$. This ideal experiment thus allows one to measure
the equation of state for temperatures larger than $T_f$. 
Based on a similar picture, Van Hove~\cite{VanHove:1982vk} argued that
the transition from a hadronic gas to a quark-gluon plasma should
result in a flattening of the mean transverse momentum $\langle
p_T\rangle$ as a function of the multiplicity.
It has been recently attempted to extract an approximate equation of
state from recent $pp$ and $p\bar p$ collision data  on this
basis~\cite{Campanini:2011bj,Ghosh:2014gra}.

The little liquid produced in an ultrarelativistic nucleus-nucleus
collision has similarities with this ideal experiment if one cuts a
thin slice perpendicular to the collision axis and looks at its
evolution in the transverse plane. 
The initial transverse velocity is initially zero, and the fluid 
expands freely into the vacuum right after the collision takes place. 
The two main differences are:
\begin{itemize}
\item{The initial temperature profile is not
uniform in a box but has a non-trivial transverse structure.}
\item{The slice expands in the longitudinal direction and its energy
  decreases as a result of the work of the longitudinal
  pressure~\cite{Bjorken:1982qr}  exerted by neighboring slices: $dE=-PdV$.}
\end{itemize}
As we shall see, both effects can be taken care of by appropriately
redefining the volume $V$ and the temperature $T_0$, and replacing the
energy per particle $E/N$ with the mean transverse mass,
where the transverse mass is defined by $m_T=\sqrt{p_T^2+m^2}$. 
Eqs.~(\ref{sfinal}) and (\ref{decomposition}) are replaced with: 
\begin{eqnarray}
\label{seff}
s(T_{\rm eff})&= &a \frac{1}{R_0^3}\frac{dN}{dy}, \nonumber \\
\frac{\epsilon(T_{\rm eff})}{s(T_{\rm eff})}&=&b\langle m_T\rangle,
\end{eqnarray}
where $R_0$ is a measure of the transverse radius, which will be
defined in Sec.~\ref{s:meanmt}, $T_{\rm eff}$ is an effective 
temperature taking into account the longitudinal cooling 
($T_{\rm eff}<T_0$), and $dN/dy$ is the multiplicity per unit rapidity,
and $a$ and $b$ are dimensionless parameters whose values are
independent of the equation of state and of the colliding energy. 
Their values will be determined in Secs.~\ref{s:meanmt} using
hydrodynamic calculations, which take into account the 
longitudinal cooling and the inhomogeneity of the initial profile. 

By measuring the mean transverse mass and the multiplicity density in
a given system at different colliding energies, one obtains the
variation of $\langle m_T\rangle$ as a function of $dN/dy$. 
Neglecting the energy dependence of the transverse size $R_0$ 
(this will be justified in Sec.~\ref{s:data}), 
the slope of this curve in a log-log
plot is the ratio of pressure over energy density, $P(T_{\rm
  eff})/\epsilon(T_{\rm 
  eff})$~\cite{Ollitrault:1991xx,Bozek:2012fw,Noronha-Hostler:2015uye}. 
Using Eqs.~(\ref{seff}), one obtains
\begin{equation}
\label{poverepsilon}
\frac{d\ln \langle m_T\rangle}{d\ln dN/dy}=
\left.\frac{d\ln\epsilon-d\ln s}{d\ln s}\right|_{T_{\rm eff}}=
\left.\frac{P}{\epsilon}\right|_{T_{\rm eff}},
\end{equation} 
where we have used the thermodynamic identities $d\epsilon=Tds$ and
$\epsilon+P=Ts$. 
Note that the dependence on the unknown coefficients $a$ and $b$
cancels in this expression. 
One thus obtains a measure of the ratio $P/\epsilon$ of 
the quark-gluon matter produced in the collision from data alone. 
The entropy density $s(T_{\rm eff})$ at which this ratio is measured,
however, depends on the  coefficient $a$, which can only be obtained
through detailed hydrodynamic simulations. These will be carried out in
Sec.~\ref{s:meanmt}.

\section{Equations of state}
\label{s:eos}

\begin{figure}[tb]
\includegraphics[width=\linewidth]{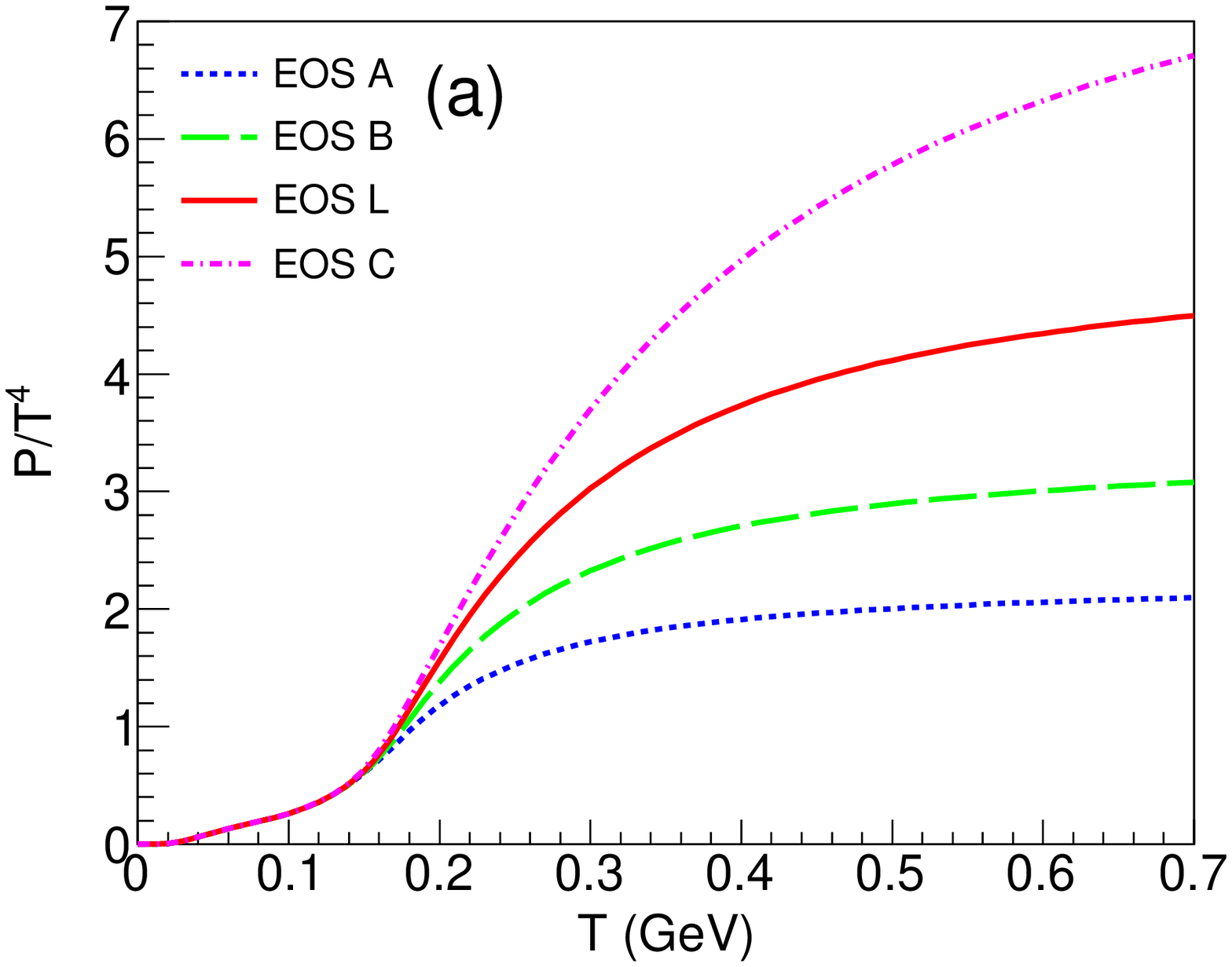}
\includegraphics[width=\linewidth]{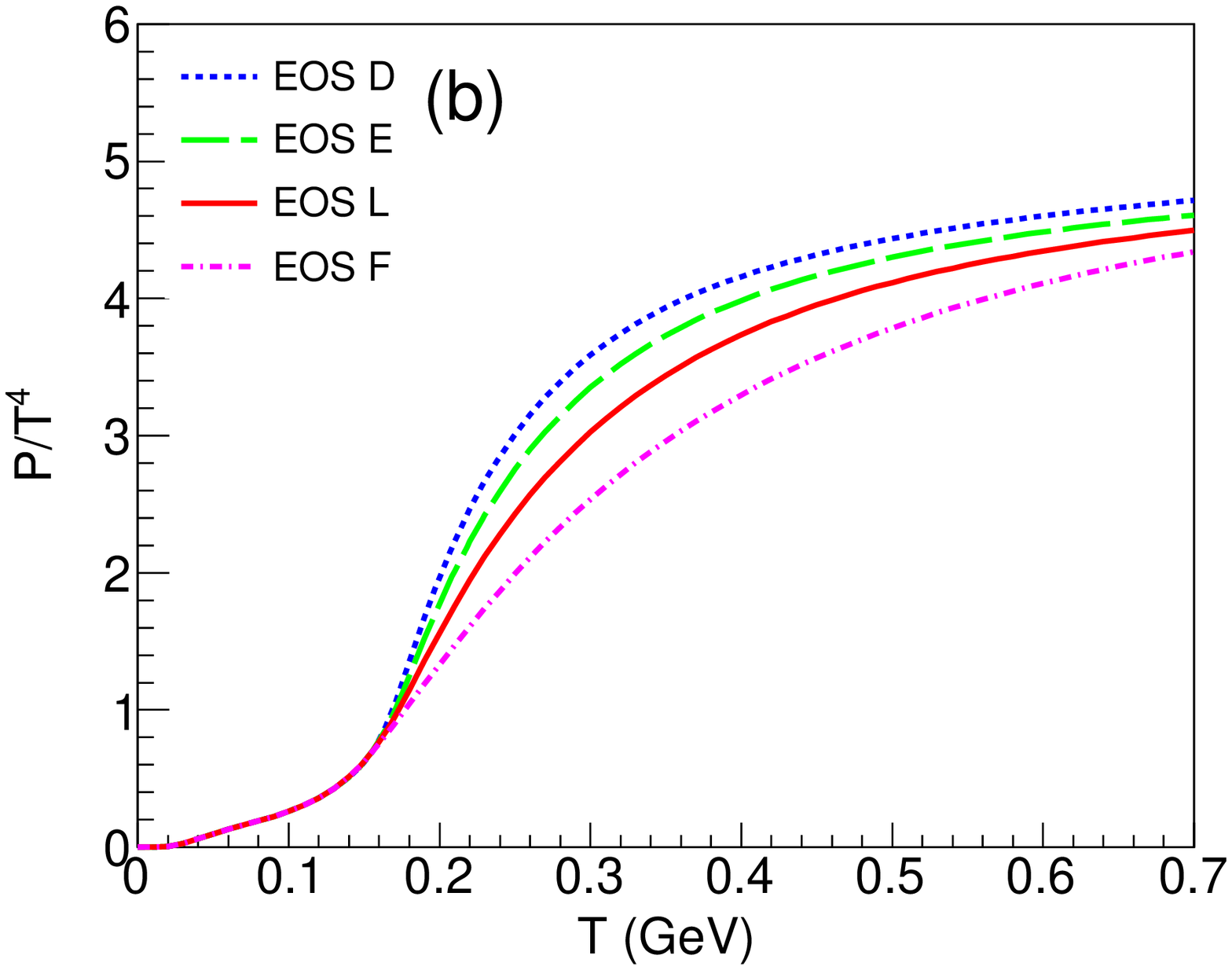}
\caption{(Color online) The pressure $P$ normalized by $T^4$ versus
  the temperature $T$.  
The curves correspond to various parameterizations obtained by varying
the number of degrees of freedom (a), or the transition temperature
(b). 
The solid line in both panels, labeled `L', corresponds to the lattice
result~\cite{Bazavov:2014pvz}. 
}
\label{fig:2}
\end{figure}

The equation of state of QCD is characterized
by a transition  from a hadronic, confined system at low temperatures
to a phase dominated by colored degrees of freedom at high
temperatures. It has been determined precisely through lattice
calculations~\cite{Borsanyi:2010cj,Bazavov:2014pvz}. 
Lattice calculations are carried out at zero baryon chemical
potential, and the matter produced at central rapidity in high-energy
collisions also has small net baryon number. We therefore choose to 
neglect net baryon density in the present study. 

In lattice calculations, one first computes the trace
anomaly $I\equiv\epsilon-3P$  as a function of the
temperature $T$, where $\epsilon$ is the energy density and $P$ the
pressure. Other quantities are then determined through the
thermodynamic relations:  
\begin{eqnarray}
\label{pressure}
\frac{P}{T^4} &=& \int_0^T \frac{I}{T^5} dT , \nonumber \\
\epsilon&=&I+3P , \nonumber \\
s&=&\frac{\epsilon+P}{T}.
\end{eqnarray}
The equation of state used in hydrodynamic calculations is
constrained, on the low-temperature side, by the condition that it
matches that of the hadron resonance gas created at
the end of the evolution~\cite{Huovinen:2009yb,Monnai:2015sca}. 
All the equations of state used in this paper match the hadron
resonance gas for temperatures smaller than 140~MeV, which is the
freeze-out temperature of our hydrodynamic calculation. 
We choose to vary the high-temperature part along two different
directions: either by varying the high-temperature limit of $P/T^4$,
which is proportional to the number of degrees of freedom of the
quark-gluon plasma (denoted as equation of state (EOS) A, B, L and C in Fig.~\ref{fig:2} (a) where EOS L corresponds to the lattice QCD-based equation of state), or by varying the
temperature range over which the transition occurs (denoted as equation of state (EOS) D, E, L and F in Fig.~\ref{fig:2}
(b)). The parameterization is explicated in
Appendix~\ref{s:eosdetails}. 
We thus span a range of equations of state around the lattice value. 
Note that the error on $P/T^4$ from lattice calculations is smaller
than $0.1$ for all $T$~\cite{Borsanyi:2010cj}. We explore a much
wider range of equations of state.

\begin{figure}[tb]
\includegraphics[width=\linewidth]{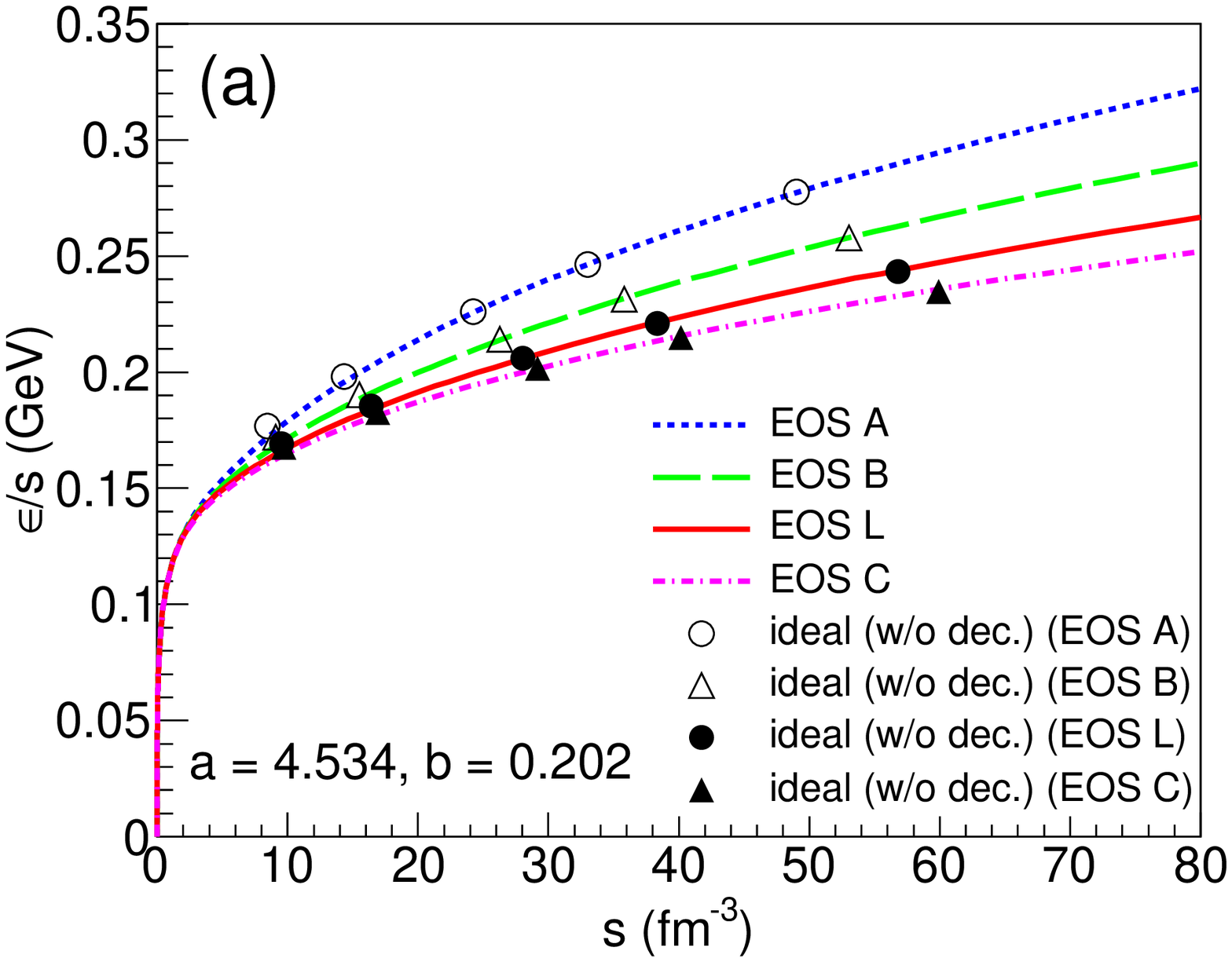}
\includegraphics[width=\linewidth]{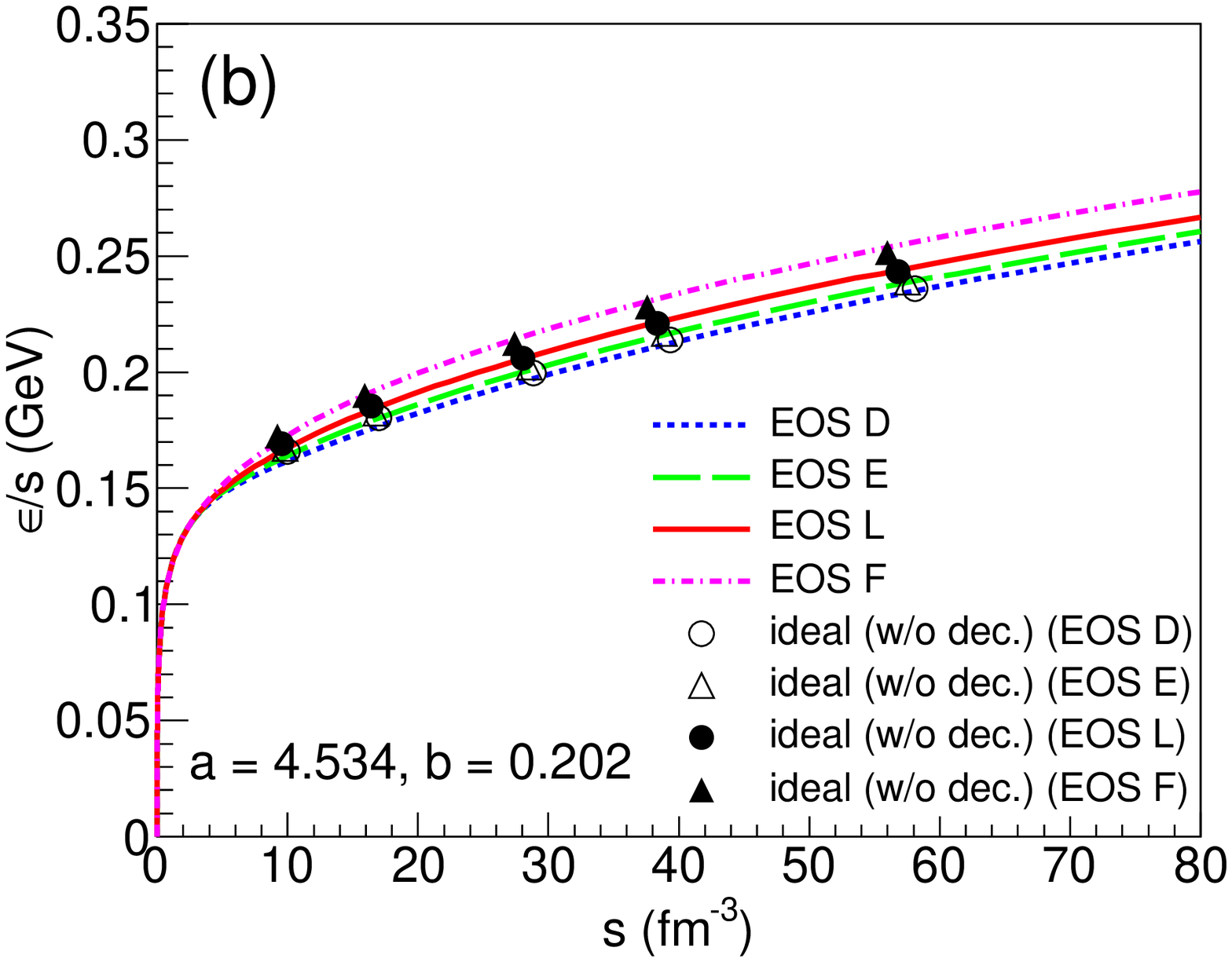}
\caption{(Color online) 
Energy over entropy versus entropy density for the equations of state
shown in Fig.~\ref{fig:2}. 
Symbols correspond to Eq.~(\ref{seff}) 
where $\langle m_T\rangle$ and $dN/dy$ are
evaluated at $T_f=140$~MeV in ideal hydrodynamics, before resonance decays. 
}
\label{fig:3}
\end{figure}

According to the picture outlined in Sec.~\ref{s:gedanken}, 
heavy-ion collisions measure the variation of the 
energy over entropy ratio as a function of the
entropy density. 
This variation is displayed in Fig.~\ref{fig:3} for the various
equations of state displayed in Fig.~\ref{fig:2}. 
Note that the ratio $\epsilon/s$ is closely related to the
temperature~\cite{VanHove:1982vk}:
\begin{equation}
\frac{3T}{4}<\frac{\epsilon}{s}<T,
\end{equation}
where the lower bound corresponds to the ideal gas limit
$P=\epsilon/3$ and the upper bound to $P=0$. 
Thus, the variation of  $\epsilon/s$ as a function of $s$ is
essentially the variation of the temperature with the entropy density. 
In the high-temperature phase, $s\propto \nu T^3$, where $\nu$ is the
effective number of degrees of freedom  of the quark gluon plasma. 
More degrees of freedom implies a smaller temperature, for the same entropy
density, which explains why the order of the curves is inverted in
Fig.~\ref{fig:3} compared to Fig.~\ref{fig:2}.

\section{Hydrodynamic calculations}
\label{s:meanmt}

In this section, we carry out hydrodynamical simulations in order to
determine the mapping between observables and the equation of state
according to Eq.~(\ref{seff}). 
We model the evolution of the fluid near midrapidity and assume
boost invariance in the longitudinal 
direction~\cite{Bjorken:1982qr}. 
We solve the transverse expansion numerically using a (2+1)-dimensional 
code~\cite{Monnai:2014kqa}. 
The initial transverse velocity is assumed to be zero  at the proper
time $\tau_0=0.4$~fm/c at which the hydrodynamic expansion starts. 
This small value of $\tau_0$ accounts for the early transverse
expansion~\cite{Ryblewski:2012rr,vanderSchee:2013pia,Keegan:2016cpi}, 
irrespective of whether or not hydrodynamics is applicable at early 
times~\cite{Vredevoogd:2008id}. 

Initial conditions are defined by the initial transverse density profile. 
The most important quantity involving initial conditions in this
study is the effective radius $R_0$ defined by: 
\begin{equation}
\label{defR}
R_0^2\equiv 2\left(\langle |{\bf x}|^2\rangle-|\langle {\bf
  x}\rangle|^2\right),
\end{equation}
where ${\bf x}$ is the position in the transverse plane, and 
angular brackets denote an average value weighted with the 
initial entropy density: 
\begin{equation}
\langle F({\bf x})\rangle \equiv
\frac{\int F({\bf x}) s({\bf x},\tau_0)d^2{\bf x} }
{\int s({\bf x},\tau_0)d^2{\bf x} }.
\end{equation}
The normalization factor 2 in Eq.~(\ref{defR}) ensures that one
recovers the correct result for a uniform entropy density profile within a
circle of radius $R_0$. 

In the ideal experiment described in Sec.~\ref{s:gedanken}, the
mapping between observables and the equation of state is independent
of the shape of the initial volume. 
For this reason, one expects that most of the dependence on the shape
of the initial density profile is through the radius $R_0$. 
This has been checked in detail in studies of transverse momentum
fluctuations~\cite{Broniowski:2009fm,Bozek:2012fw,Bozek:2017jog}, where it was shown
that the mean transverse momentum in hydrodynamics is sensitive to
initial state fluctuations only through fluctuations of $R_0$. 
We have checked it independently by comparing two standard models of 
initial conditions, the Monte Carlo Glauber model~\cite{Miller:2007ri} and the
MCKLN~\cite{Drescher:2006ca} model, as will be explained below. 
The default setup of our hydrodynamic calculation uses a Monte Carlo
Glauber simulation of 0-5\% most central Au+Au collisions 
where the energy density is a sum of contributions of
binary collisions, and the contribution of each collision is a
Gaussian of width 0.4~fm centered half way between the colliding
nucleons. The resulting density profile is centered, and then averaged over a
large number of events in order to obtain a smooth profile~\cite{Qiu:2011iv}. 
The normalization of the density profile determines the multiplicity
$dN/dy$. We run each calculation with 5 different normalizations
spanning a range which covers the LHC and RHIC data which will be used in Sec.~\ref{s:data}.

\subsection{Ideal hydrodynamics}
\label{s:ideal}

We first carry out ideal hydrodynamic simulations 
for all the equations of
state displayed in Fig.~\ref{fig:2}. 
The fluid is converted into hadrons through the standard Cooper-Frye
freeze-out procedure~\cite{Cooper:1974mv} at a temperature $T_f=140$~MeV. 
We include all hadron resonances with $M<2.25$~GeV, 
and compute $\langle m_T\rangle$ and $dN/dy$ directly at freeze-out,
before resonances decay. Our goal here is to mimic as closely as 
possible the ideal experiment outlined in Sec.~\ref{s:gedanken}. 

The symbols in Fig.~\ref{fig:3} correspond to the right-hand side 
of Eq.~(\ref{seff}), where the
dimensionless parameters $a$ and $b$ have been fitted 
to achieve the best possible agreement with the 
left-hand side. 
There are 5 points for each equation of state, which
correspond to different initial temperatures. 
The overall agreement is excellent, and shows that the variation of 
$\langle m_T\rangle$ as a function of $(1/R_0^3)(dN/dy)$ is determined
by the equation of state. 

In order to test that this mapping is independent of initial
conditions, we have carried out a calculation with MCKLN initial
conditions. 
While both models give values of $R_0$ that differ by 5\%, they yield
the same value of $\langle m_T\rangle$ when compared at the same value
of $(1/R_0^3)dN/dy$.

Let us now comment on the order of magnitude of the fit parameters $a$
and $b$. 
First, compare Eq.~(\ref{decomposition}) and the second line of
Eq.~(\ref{seff}). 
The entropy per particle at freeze-out before decays is
$(S/N)_{T_f}=6.5$ in this calculation.
The transverse mass of a particle is smaller than its energy, since it
does not include the longitudinal momentum $p_z$. 
The relevant longitudinal momentum here is that relative to the fluid,
which cannot be measured, since data are integrated over all 
fluid rapidities. 
The value of $b=0.202$  is slightly larger than $(N/S)_{T_f}=0.154$, 
and thus compensates for the loss of longitudinal momentum. 

We now discuss the order of magnitude of $a$. The main difference
between the ideal experiment described in Sec.~\ref{s:gedanken} and
the real experiment is that the energy of the fluid slice decreases 
as a result of the work done by the longitudinal pressure. 
In ideal hydrodynamics, this cooling is only significant at 
early times: After the transverse expansion sets in, the pressure
decreases very rapidly, the work becomes negligible
and the energy stays constant. 
A rough, but qualitatively correct, picture is that the expansion is
purely longitudinal during a time $\tau_{\rm eff}$ and that the energy is
conserved for $\tau>\tau_{\rm eff}$~\cite{Ollitrault:1991xx}. 
For dimensional reasons, $\tau_{\rm eff}=fR_0$, where $f$ is of order
unity. 
The volume at $\tau_{\rm eff}$ is $V=\pi R_0^2 \tau_{\rm eff}=\pi fR_0^3$.
Inserting this value into Eq.~(\ref{sfinal}) and identifying the
right-hand side with the first line of Eq.~(\ref{seff}), one obtains
$f\simeq 0.5$, in agreement with the value obtained in previous
calculations~\cite{Ollitrault:1991xx}. 
Ideal hydrodynamics thus probes the equation of state at a time
$\tau_{\rm eff}\sim 0.5R_0$, which is the typical time at which transverse 
flow and elliptic flow develop~\cite{Kolb:1999es,Kolb:2000sd,Gombeaud:2007ub}.

\subsection{Resonance decays}
\label{s:decays}

\begin{figure}[tb]
\includegraphics[width=\linewidth]{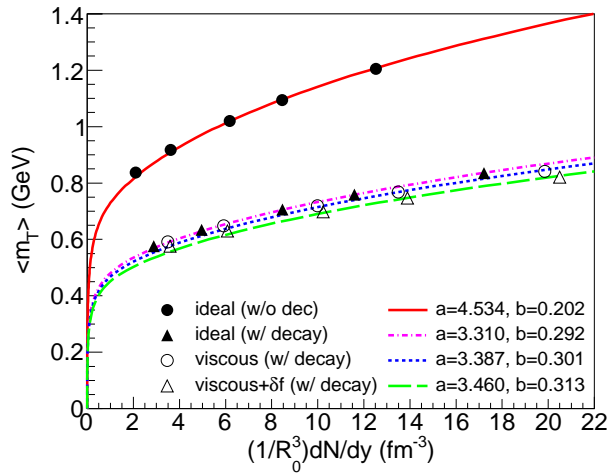}
\caption{(Color online)
Comparison of results from various hydrodynamic calculations. From top
to bottom: ideal hydrodynamics before and after decays, 
viscous hydrodynamics without and with viscous correction at
freeze-out. 
}
\label{fig:decays}
\end{figure}
The largest correction to the naive ideal fluid picture comes from
decays occurring through strong or electromagnetic interactions, 
which occur after freeze-out, but before the 
daughter particles reach the 
detectors.
We compute particle spectra after strong and 
electromagnetic decays, but before weak decays.
Decays are treated in Ref.~\cite{Sollfrank:1990qz}, by assuming
that the decay rate is proportional to the invariant phase space. 
After decays, the only remaining particles are pions, kaons, nucleons
and strange baryons. 
In this preliminary study, we neglect strange baryons, which are a
small fraction of the total number of particles, and are 
identified in separate analyses~\cite{Abelev:2013xaa}. 
We therefore evaluate the multiplicity $dN/dy$ and the mean transverse
mass including only pions, kaons, and (anti)nucleons, both charged and
neutral. 
As shown in Fig.~\ref{fig:decays}, 
decays increase the multiplicity by 40\%. 
They also conserve the total energy, so that $\langle m_T\rangle$
decreases, while the product $\langle m_T\rangle dN/dy$ only changes 
by a few percent.  

\begin{figure}[tb]
\includegraphics[width=\linewidth]{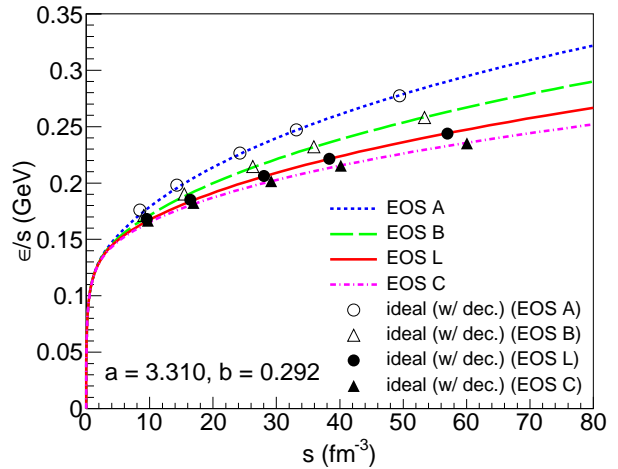}
\caption{(Color online) 
Same as Fig.~\ref{fig:3} (a) after resonance decays. 
}
\label{fig:3bis}
\end{figure}

Since the increase of $dN/dy$ due to decays depends solely on the
freeze-out temperature, but is independent of the colliding energy and
the equation of state, 
decays amount to further rescalings of $\langle m_T\rangle$ and 
$dN/dy$. They can be taken into account by modifying the values of
the coefficients $a$ and $b$ in Eq.~(\ref{seff}).
We again determine the values of $a$ and $b$ through a simultaneous
least-square fit to all equations of state. The result is shown in 
Fig.~\ref{fig:3bis}, where only the equations of state of
Fig.~\ref{fig:1} (a) are shown. 
After rescaling, the effective entropy density of the fluid is
unchanged: locations of symbols in Fig.~\ref{fig:3} (a) and
Fig.~\ref{fig:3bis} are identical to within less than 0.5\%. 
The fact that they are identical confirms that Eqs.~(\ref{seff})
reconstruct thermodynamic properties of the fluid. 

A more realistic description of the hadronic stage should include not
only decays, but also rescatterings, for instance by coupling
hydrodynamics to a transport
code~\cite{Teaney:2000cw,Petersen:2008dd,Song:2010aq}. 
It has been recently shown~\cite{Ryu:2017qzn} that transverse momentum
spectra are remarkably independent of the temperature at which one
switches from the hydrodynamic to the transport description, which
implies that our results would be unchanged if we switched from a
hydrodynamic description to a transport calculation at a temperature
larger than 140~MeV. Below 140~MeV, effects of hadronic scatterings
are suppressed due to the lower density. Our choice of $T_f$ allows us
to roughly reproduce observed particle ratios, in agreement with 
Ref.~\cite{Ryu:2017qzn}. This is important as the mean $m_T$, averaged
over all particle species, strongly depends on particle ratios.

\subsection{Viscosity}
\label{s:corrections}

We finally study viscous corrections to the ideal fluid picture.
We use ``minimal'' shear viscosity $\eta/s=1/4\pi$~\cite{Kovtun:2004de}
and bulk viscosity $\zeta/s = 2 (1/3 - c_s^2) \eta/s$~\cite{Buchel:2007mf}
based on the gauge-string correspondence, where
$c_s$ is the sound velocity. The relaxation times are also
conjectured in the holographic approach \cite{Natsuume:2007ty}.
Viscosity modifies the equations of motion of the
fluid~\cite{Israel:1979wp}, 
and the momentum distribution of particles at
freeze-out~\cite{Teaney:2003kp,Monnai:2009ad}. 
We show both effects separately in Fig.~\ref{fig:decays}.
The main effect of viscosity is to 
increase the multiplicity for a given initial condition, which is
a consequence of the entropy increase due to dissipative processes. 
On the other hand, the value of $\langle m_T\rangle$ changes little, 
which is due to a partial cancellation between effects of shear
viscosity (which increases $\langle m_T\rangle$) and bulk viscosity 
(which decreases $\langle m_T\rangle$) \cite{Monnai:2009ad}.

\begin{figure}[tb]
\includegraphics[width=\linewidth]{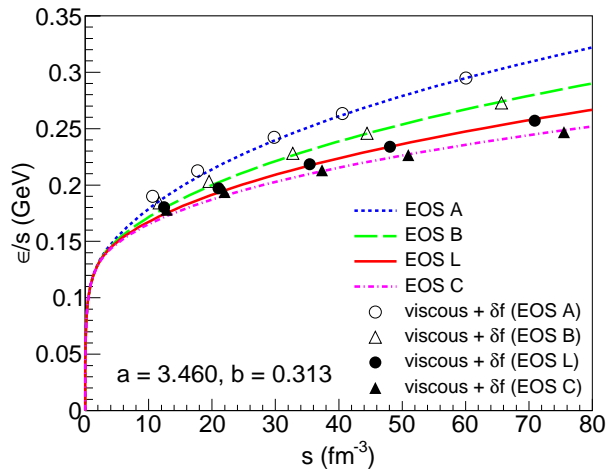}
\caption{(Color online) 
Same as Fig.~\ref{fig:3bis} with shear and bulk viscosity included. 
}
\label{fig:eoshydro}
\end{figure}
The values of $\langle m_T\rangle$ and $dN/dy$ can again be matched to
the equation of state through Eqs.~(\ref{seff}). 
We again determine the values of $a$ and $b$ which give the best
simultaneous fit to all equations of state. 
The result is displayed in Fig.~\ref{fig:eoshydro}, where only the
equations of state of Fig.~\ref{fig:1} (a) are shown. 
This figure shows that viscous corrections do not alter qualitatively
the ideal fluid picture, and that the variation of the mean
transverse mass with the multiplicity density is still driven by the
equation of state in the presence of viscosity. 
Comparison with Fig.~\ref{fig:3bis} shows that symbols are shifted to
the right, which means that for the same initial temperature, viscous
hydrodynamics results in a higher effective entropy density. 
The reason is that entropy is produced in the off-equilibrium
processes. 

The actual value of the shear and bulk viscosity are not known
precisely. Since $a$ and $b$ depend slightly on the viscosity, 
the uncertainty on the viscosity translate into an uncertainty on the
mapping of experimental data onto the equation of state through
Eq.~(\ref{seff}). 
The upper bound on constant $\eta/s$ from heavy-ion data is typically
$0.2$~\cite{Shen:2015msa}. It has been recently noted that the
inclusion of bulk viscosity tends to lower the preferred value 
of the shear viscosity~\cite{Ryu:2017qzn}, so that $\eta/s<0.2$ seems
conservative. 
We assume that viscous corrections are proportional to the viscosity, 
therefore the uncertainty can be inferred from the difference between
our viscous and ideal calculations. 
The uncertainty on $a$ is $7\%$ and amounts on an uncertainty on
the effective entropy density $s_{\rm eff}$. 
The uncertainty on $b$ is $11\%$ and is essentially an uncertainty on the
corresponding temperature. 
Note, however, that the dependence on $a$ and $b$ cancels in the
logarithmic slope, Eq.~(\ref{poverepsilon}), and the ratio
$P/\epsilon$ can be determined precisely even if transport
coefficients are not precisely determined. 

\section{Comparison with data}
\label{s:data}

\begin{table}
\begin{tabular}{c|c|c|c|c|c}
\hline
\hline
$\sqrt{s}$&
$dN/dy$&$\langle m_T\rangle$&$R_0$ &

$s_{\rm eff}$& $T_{\rm eff}$\cr
(GeV)&
&(MeV)&(fm)&
(fm$^{-3}$)& 
(MeV)\cr
\hline
5020~\phantom{\cite{Abelev:2013vea}}
& ?&?&$6.21$
& $48.1 \pm 3.1 $
& $292\pm 5$
\cr
2760~\cite{Abelev:2013vea}
&$2764\pm 177$ &$686\pm 19$&$6.17$
& $40.7\pm 2.6$
&$279\pm 5$
\cr
\phantom{1}200~\cite{Adler:2003cb}
&$1146\pm 79\phantom{1}$ &$589\pm 33$&$5.97$
&$18.6\pm 1.3$
&$227\pm 4$
\cr
\phantom{1}200~\cite{Abelev:2008ab}
&$1220\pm 97\phantom{1}$ &$590\pm 48$&$5.97$
&$19.9\pm 1.6$
&$231\pm 5$
\cr
\phantom{1}130~\cite{Adams:2003xp} 
&$1042\pm 77\phantom{1}$ &$560\pm 41$&$5.93$
&$17.1\pm 1.3$
&$223 \pm 4$
\cr
62.4~\cite{Abelev:2008ab}
&$\phantom{1}867\pm 65\phantom{1}$ &$549\pm 28$&$5.92$
&$14.3\pm 1.1$
&$214 \pm 4$
\cr
\hline
\hline
\end{tabular}
\caption{\label{tableseff} 
Results for Pb+Pb collisions at the
LHC and Au+Au collisions at RHIC. The centrality is 0-6\%  for 130~GeV
data and 0-5\% for all other energies. 
The first columns give our Values of $\langle m_T\rangle$ and $dN/dy$,
obtained by extrapolating the measurements (see text). 
The 3rd column is the value of $R_0$ we use in Eq.~(\ref{seff}), which
is obtained from a Glauber model, but subject to significant
theoretical uncertainty (see text). 
The last columns give the values of the effective entropy density defined by
Eq.~(\ref{seff}), and of the corresponding temperature if the
equation of state is taken from lattice QCD. 
Error bars on $s_{\rm eff}$ and $T_{\rm eff}$ are experimental only. 
}
\end{table}

We now discuss to what extent existing data constrain the equation of
state. 
Both $dN/dy$ and $\langle m_T\rangle$ require spectra of pions, kaons
and protons. 
Such data have been published by STAR~\cite{Abelev:2008ab} and
PHENIX~\cite{Adler:2003cb} at the Relativistic Heavy Ion Collider
(RHIC) and by ALICE~\cite{Abelev:2013vea} at the Large Hadron Collider (LHC).
PHENIX and ALICE data for protons are corrected for the contamination
from weak $\Lambda$ decays, while STAR data are not. 
We correct STAR data by assuming that a fraction $35\%\pm 10\%$ of
protons come from $\Lambda$ decays, as determined by the PHENIX
analysis~\cite{Adler:2003cb}. 
Particles are only identified within a limited $p_T$ range, which
depends on the experiment, and spectra must be extrapolated in order
to obtain $dN/dy$ and $\langle m_T\rangle$. 
These extrapolations are discussed in Appendix~\ref{s:bwfits}. 
The data we use are for charged particles, and we need $\langle
m_T\rangle$  and $dN/dy$ for all hadrons, including neutral ones. 
Yields of neutral particles are obtained assuming isospin symmetry. 
The resulting values of $\langle m_T\rangle$ and $dN/dy$ are given in
Table~\ref{tableseff}. 
For 200~GeV, we include both STAR and PHENIX measurements, which are
slightly different, but compatible within errors. 

In order to convert the multiplicity $dN/dy$ into a density, one needs
an estimate of the initial transverse size $R_0$. 
This quantity, which represents the mean square radius of the initial
density profile, is not measured and can only be estimated in a
model. As we shall see, it turns out to be the largest source of uncertainty
when constraining the equation of state from data. 
In particular, the uncertainty from $R_0$ is larger than the
uncertainty from transport coefficients. 

We discuss how we estimate $R_0$. 
Note that the transverse size fluctuates event to event, even in a narrow
centrality window~\cite{Broniowski:2009fm}.
Ideally, we would like to estimate the average value over events of
$(1/R_0^3)dN/dy$. Since the input available from experiment is an average
of $dN/dy$, for the sake of simplicity, we estimate the average value of
$R_0$ over many events to divide $dN/dy$ for our analyses.
We use the same Monte Carlo Glauber model as in our hydrodynamic
calculation. 
The resulting values, averaged over many events, are given in
Table~\ref{tableseff}. The 
MCKLN model~\cite{Drescher:2006ca} gives values $5\%$ smaller, which
implies that the density is $15\%$ larger. 
This shows that the uncertainty on the transverse size is significant.

However, the variation of $R_0$ with colliding energy for a given
system is small, so that the evolution of the density is mostly
driven by the increase in the multiplicity $dN/dy$. 
Therefore, uncertainties on $R_0$ cancel 
when comparing two different collision energies. 
The variation of the mean transverse mass with $dN/dy$ directly gives
the ratio $P/\epsilon$, as shown by Eq.~(\ref{poverepsilon}). 
As pointed out in Sec.~\ref{s:corrections}, 
uncertainties from the viscosity also cancel in this 
energy dependence. Using PHENIX and ALICE data, which span a wide
range of $dN/dy$, and taking into account the different sizes of Au
and Pb nuclei, Eq.~(\ref{poverepsilon}) gives
\begin{equation}
\label{ourpovereps}
\left.\frac{P}{\epsilon}\right|_{T_{\rm eff}}=0.21\pm 0.10,
\end{equation}
where the error is solely from experiment. 

The only significant theoretical uncertainty is on the effective
temperature $T_{\rm eff}$ at which this ratio is measured.  
We provide in Table~\ref{tableseff} the values of the effective
entropy density $s_{\rm eff}$ given by Eq.~(\ref{seff}), where $a$ is
given by our viscous hydrodynamic calculation. 
The value at $5.02$~TeV, where identified particle spectra are not yet
published, is obtained by assuming that the relative
increase in $dN/dy$ from $2.76$~TeV equals that of
$dN_\mathrm{ch}/d\eta$, that is, 20\%~\cite{Adam:2015ptt}. 
As discussed in Sec.~\ref{s:corrections}, the uncertainty on $s_{\rm
  eff}$ from transport coefficients is $7\%$, and that 
from the transverse size $R_0$ is at least 15\%. 

The value of the temperature $T_{\rm eff}$ corresponding to $s_{\rm
  eff}$  can only be obtained if the equation of state is known. 
The values in the last column of Table~\ref{tableseff} correspond to
the lattice equation of state. 
Lattice calculations give $P/\epsilon=0.23$ for a temperature half-way
between the values of $T_{\rm eff}$ corresponding to 200~GeV and
2.76~TeV. The experimental value, Eq.~(\ref{ourpovereps}), is
compatible with the lattice result. 
Experiments at $\sqrt{s}=5.02$~TeV, for which identified particle
spectra are yet unpublished, will probe the equation of state at a
temperature close to 300~MeV. 
Note that the theoretical uncertainty of $\simeq 20\%$ on $s_{\rm eff}$
translates into an uncertainty  $\sim 15$~MeV on the effective
temperature at the LHC, which is dominated by the uncertainty on the
initial transverse radius $R_0$.

\begin{figure}[tb]
\includegraphics[width=\linewidth]{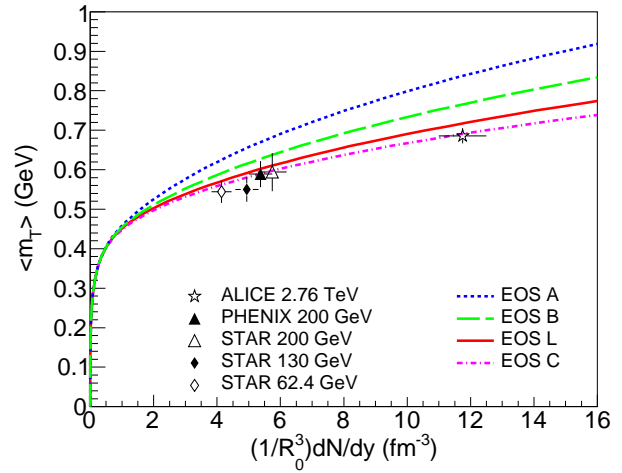}
\caption{(Color online) 
Experimental data from Table~\ref{tableseff}, compared to value given
by various equations of state and Eqs.~(\ref{seff}), where $a$ and $b$
are taken from our viscous hydrodynamic calculation, and $R_0$ from a
Glauber model (see text). 
}
\label{fig:data}
\end{figure}
Figure~\ref{fig:data} shows the comparison between experimental data
and the values obtained from the equation of state through
Eqs.~(\ref{seff}), where $a$ and $b$ are taken from our viscous
hydrodynamic calculation (see Fig.~\ref{fig:eoshydro}). 
With the minimal viscosity chosen in this calculation, LHC data
slightly favor the equation of state C, which has a larger pressure
than the lattice equation of state. With a higher viscosity, however,
the lattice equation of state would be preferred. 
Equations of state A and B are ruled out: as already well known,
heavy-ion data favor a soft equation of state. 
Note that current experiments only probe the equation of
state up to $T\sim 300$~MeV (see Table~\ref{tableseff}).

\section{Conclusions}
\label{s:conclusions}

We have shown that in central nucleus-nucleus collisions, the
variation of the mean 
transverse mass as a function of the multiplicity density is, up to
rescaling factors, driven by the variation of the energy over entropy
ratio $\epsilon/s$ as a function of the entropy density $s$. 
Each collision energy probes the equation of state at a different
entropy density $s_{\rm eff}$, which corresponds 
roughly to the average density at a time 
$\tau_{\rm eff}\sim 3$~fm/$c$. RHIC and LHC experiments 
probe the equation of state for temperatures up to $\sim 300$~MeV.

The largest source of uncertainty at the theoretical level is the initial
transverse size $R_0$. 
The uncertainty from unknown transport coefficients (shear and bulk
viscosity) is twice smaller. 
These theoretical uncertainties cancel if one measures the {\it
  evolution\/} of the mean 
transverse mass as a function of collision energy, which gives direct
access to the pressure over energy density ratio $P/\epsilon$ of the
quark-gluon plasma. 

This analysis requires precise experimental data on identified
particle spectra. 
One could think of replacing the transverse mass with the transverse
momentum, and the rapidity by the pseudorapidity, which was the
original idea of Van Hove~\cite{VanHove:1982vk}, and would allow to
work with unidentified particles. However, we have checked that the
mapping onto the equation of state is not as good in this case. 

The  value of $P/\epsilon$ obtained from the evolution of spectra from
RHIC to LHC energies is compatible with the lattice equation of 
state, but with large errors. Carrying out an energy scan at the LHC
with a single detector would greatly improve the quality of the
measurement. 


\section*{Acknowledgements}
We thank Matt Luzum and Michele Floris for discussions and Jean-Paul
Blaizot for useful comments on the manuscript. 
AM is supported by JSPS Overseas Research Fellowships.

\appendix
\section{Varying the equation of state} 
\label{s:eosdetails}

The equation of state is constructed by connecting the trace 
anomaly of the hadron resonance gas model smoothly to 
that of lattice QCD~\cite{Bazavov:2014pvz}. To systematically 
generate variations of the equation of state, modification 
is made through two factors $c_w$ and $c_h$ in the QGP 
phase for our analyses. The expression reads:
\begin{eqnarray}
I(T) &=& \frac{1}{2} \bigg[ 1 - \tanh \bigg(\frac{T-T_s}{\Delta T_s} \bigg) \bigg] I_{\mathrm{HRG}}(T) \nonumber \\
&+& \frac{c_h}{2} \bigg[ 1 + \tanh \bigg(\frac{T-T_s}{\Delta T_s} \bigg) \bigg] I_{\mathrm{lat}}(T_w),
\end{eqnarray}
where $T_w = T_s + c_w(T-T_s)$. $c_w$ and $c_h$ are 
associated with the width and the magnitude of $I(T)$ in the 
QGP phase, respectively. $c_w = 1$ and $c_h = 1$ 
recover the lattice QCD result. The hadronic equation of state 
is left untouched because, as mentioned earlier, the 
Cooper-Frye formula requires that kinetic theory reproduces 
the equation of state used in the hydrodynamic model at 
freeze-out for energy-momentum conservation. When one 
chooses $T_s = 160$~MeV and $\Delta T_s = 0.1 T_s$, 
this is satisfied at and below $T = 140$ GeV.

\begin{figure}[tb]
\includegraphics[width=.9\linewidth]{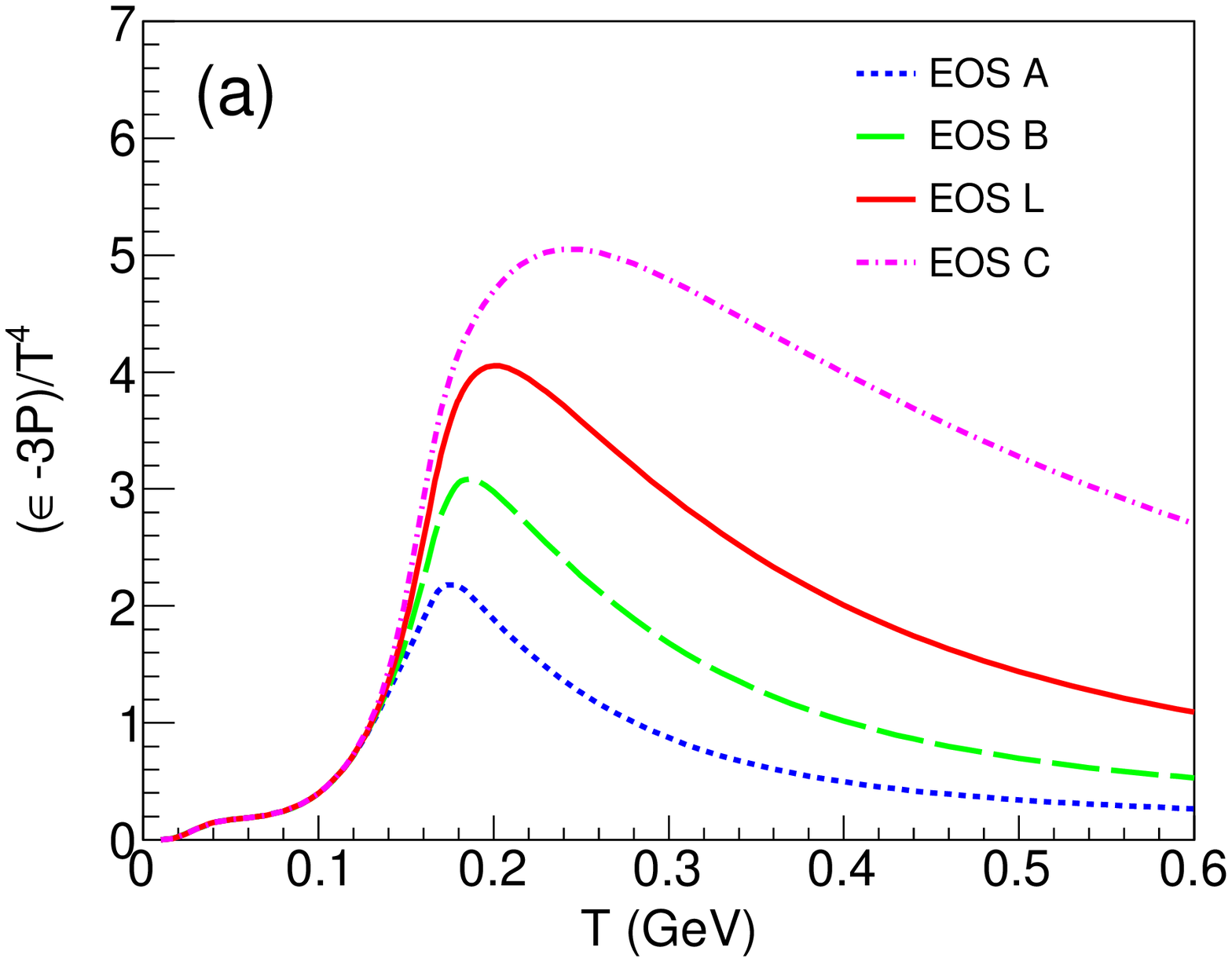}
\includegraphics[width=.9\linewidth]{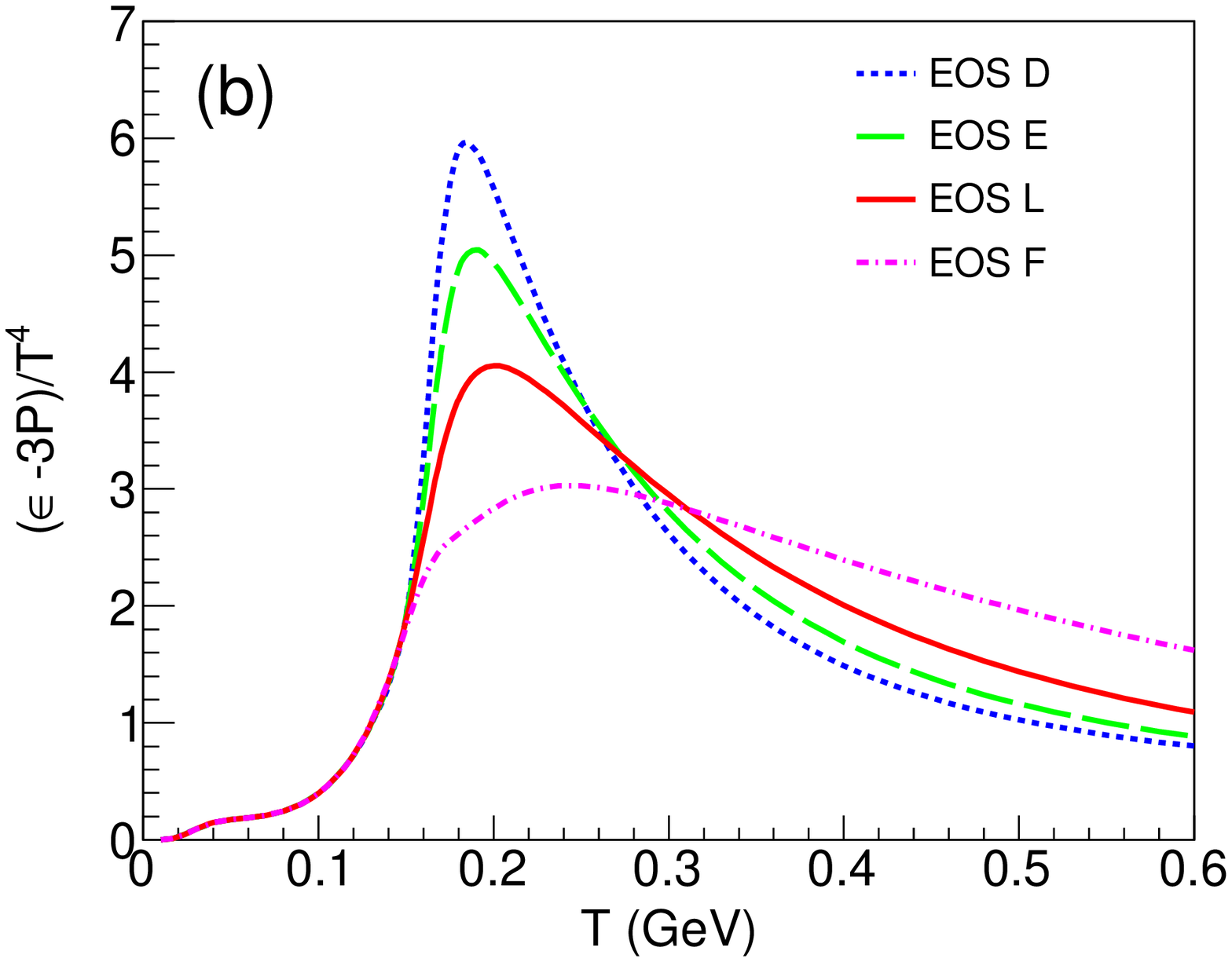}
\caption{(Color online) The trace anomaly normalized by $T^4$ 
versus the temperature $T$. The curves correspond to various 
parameterizations obtained by varying the number of 
degrees of freedom (a), or the transition temperature (b).}
\label{fig:1}
\end{figure}

The pressure is obtained through the thermodynamic relations
(\ref{pressure}). Since the trace anomaly is integrated, $c_w$ 
and $c_h$ have to be modified simultaneously to shift the pseudo-critical 
temperature and change the effective number of degrees 
of freedom in the pressure or the entropy density (Fig.\ref{fig:1}).

We first consider a set of equation of state with different numbers 
of QGP degrees of freedom by choosing $(c_w,c_h)$ = 
$(2,0.5)$, $(1.5, 0.75)$, $(1,1)$, and $(0.5,1.25)$. They are labeled 
as EOS A, B, L and C, respectively. The normalized pressure 
as a function of the temperature for each equation of state is plotted 
in Fig.~\ref{fig:2} (a). It is note-worthy that we consider an 
equation of state which exceeds the Stefan-Boltzmann limit 
with the last parameter set $(0.5,1.25)$. We also vary the 
pseudo-critical temperature by setting the parameters to 
$(c_w,c_h)$ = $(2,1.5)$, $(1.5, 1.25)$, $(1,1)$, and $(0.5,0.75)$ as 
shown in Fig.~\ref{fig:2} (b), which are labeled as EOS D, E, L and F. 
The equation of state becomes harder for larger $T_c$ 
because it is fixed on the hadronic side.

\section{Identified particle spectra at RHIC and LHC} 
\label{s:bwfits}

\begin{figure}[tb]
\includegraphics[width=.49\linewidth]{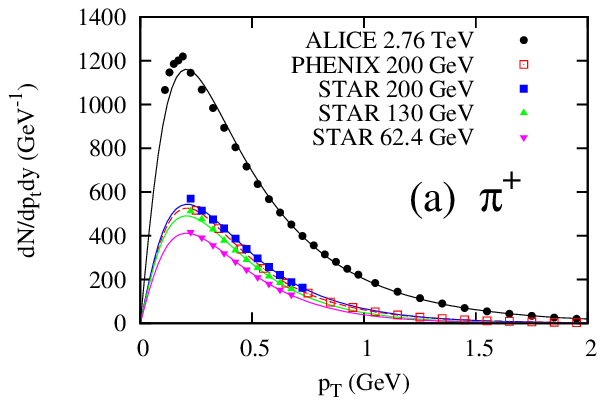}
\includegraphics[width=.49\linewidth]{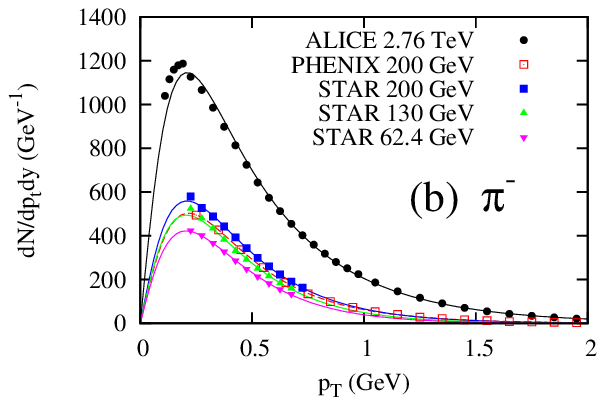}
\includegraphics[width=.49\linewidth]{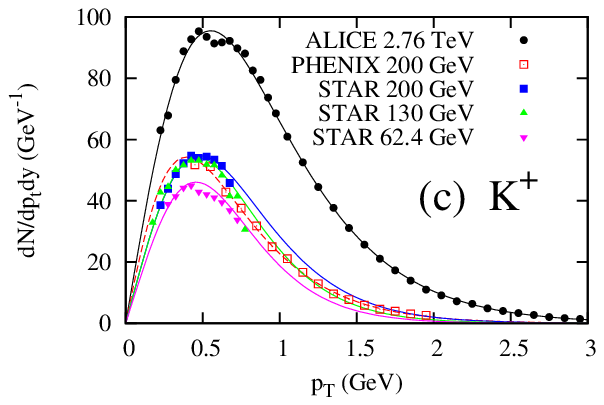}
\includegraphics[width=.49\linewidth]{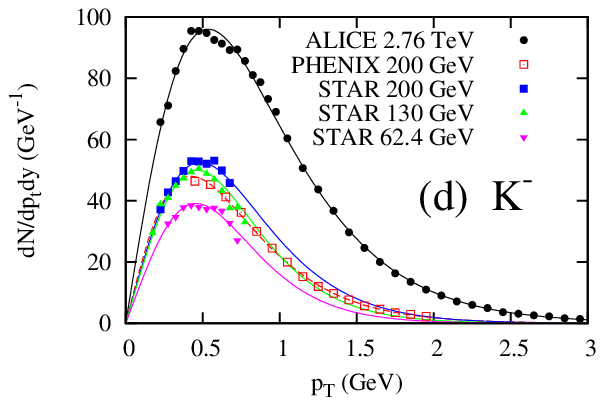}
\includegraphics[width=.49\linewidth]{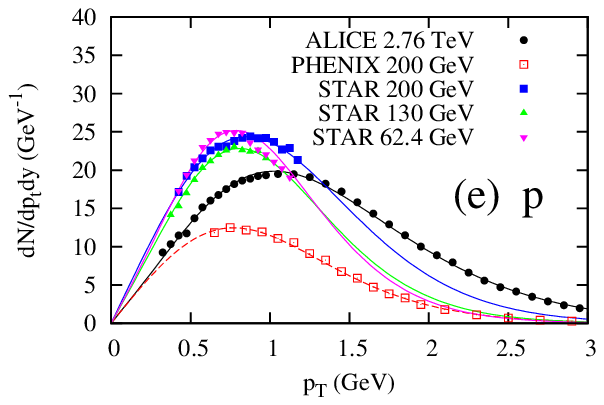}
\includegraphics[width=.49\linewidth]{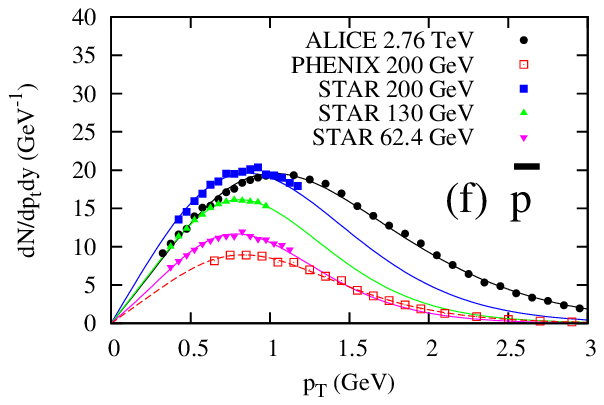}
\caption{(Color online) Transverse momentum distributions of
  identified particles in Pb+Pb collisions at the LHC and Au+Au
  collisions at RHIC. The centrality range is 0-5\% for all sets of
  data except 130~GeV data which are 0-6\%. 
  Symbols are data from ALICE~\cite{Abelev:2013vea},
  PHENIX~\cite{Adler:2003cb} and
  STAR~\cite{Adams:2003xp,Abelev:2008ab}. 
  Solid lines are blast-wave fits (see text). 
Each panel corresponds to a different particle species: positive (a)
and negative (b) pions, positive (c) and negative (d) kaons, protons
(e) and antiprotons (f). 
STAR data for protons and antiprotons also include secondary products of
$\Lambda$ and $\bar\Lambda$ decays, which explain the larger values. 
Experimental errors are not shown for sake of readability, but they
are taken into account in the fits. 
}
\label{fig:expdata}
\end{figure}

In order to estimate the mean transverse mass per particle from
experimental data, we use as input $p_T$ spectra of 
identified charged hadrons in the central rapidity region. 
More specifically, we use data for charged pions, charged kaons,
protons and antiprotons, which are shown as symbols in
Fig.~\ref{fig:expdata}.  
These plots show the probability distribution of $p_T$ near
midrapidity, $dN/dp_Tdy$. 
Experimental data are shown as symbols. 
Pion and kaon yields increase smoothly with
collision energy as expected. 
This does not appear to hold for proton and antiprotons, but the
reason is simply that STAR data for protons and antiprotons include,
in addition to primary particles, secondary products of weak $\Lambda$
and $\bar\Lambda$ decays. 
Apart from this difference, PHENIX and STAR data at 200~GeV are
compatible within error bars.  

The effect of the net baryon number becomes visible at the lower
energies:  it results in more protons than antiprotons at midrapidity,
and also slightly more $K^+$ than $K^-$ because 
the strangeness chemical potential is non-vanishing in 
the presence of the net baryon chemical potential $\mu_B$ owing to 
the strangeness neutrality condition. 
While the differences between particles and antiparticles are linear
in $\mu_B$, the total multiplicities
are even functions of $\mu_B$, hence effects of net baryon number only
appear to order $\mu_B^2$. We assume that they are negligible down to
62.4~GeV.

Particles are identified only in a limited $p_T$ range which depends on
the experiment. In order to evaluate the mean $m_T$, we need to 
extrapolate the measured spectrum to the whole $p_T$ range. 
These extrapolations are done with blast-wave fits~\cite{Schnedermann:1993ws}. 
For ALICE data, we fit each particle species 
independently, as in the experimental paper~\cite{Abelev:2013vea}. 
The resulting values of $dN/dy$ and $\langle p_T\rangle$ are given in
Tables~\ref{tabledndy} and \ref{tablemeanpt}. They are very close to
the values in the experimental paper. The small differences, which are
much smaller than 
error bars, can be ascribed to different fitting algorithms. 
For sake of consistency, we also use blast-wave fits to extrapolate PHENIX
data~\cite{Adler:2003cb}. 
The resulting values of $dN/dy$ and $\langle
p_T\rangle$ differ somewhat from the experimental values which use a
different extrapolation scheme, but are compatible within error bars. 
For STAR data, the $p_T$ range is too limited to fit each particle
species independently: therefore, we follow the recommendation of the
experimental paper~\cite{Abelev:2008ab} and carry out a simultaneous
fit for kaons and (anti)protons. 
For pions, however, we carry out an independent blast-wave fit as for 
PHENIX data. 
Agreement between STAR and PHENIX pion yields at 200~GeV is much
better than in the corresponding experimental papers, which suggests
that the differences were mostly due to the different extrapolation 
methods. 

Finally, the values of $\langle m_T\rangle$, which are needed in this
paper, are listed in Table~\ref{tablemeanmt}. 

\onecolumngrid

\begin{table}
\begin{tabular}{c|c|llllllllllll}
\hline
\hline
exp.&$\sqrt{s}$~[GeV]&$\pi^+$&&$\pi^-$&&$K^+$&&$K^-$&&$p$&&$\bar p$\cr
\hline
ALICE & 2760
&732.3 &{\it 733$\pm$54}
&731.0 &{\it 732$\pm$52}
&109.0 &{\it 109$\pm$9}
&108.6 &{\it 109$\pm$9}
&33.6 &{\it 34$\pm$3}
&33.2 &{\it 33$\pm$3}
\cr
PHENIX& 200
&306.5 &{\it 286.4$\pm$24.2}
&297.6 &{\it 281.8$\pm$22.8}
&48.1 &{\it 48.9$\pm$6.3}
&43.8 &{\it 45.7$\pm$5.2}
&16.4 &{\it 18.4$\pm$2.6}
&11.7 &{\it 13.5$\pm$1.8}
\cr
STAR& 200
&310.6 &{\it 322$\pm$25}
&315.1  &{\it 327$\pm$25}
&51.3 &{\it 51.3$\pm$6.5}	
&49.2 &{\it 49.6$\pm$6.2}	
&34.5 &{\it 34.7$\pm$4.4}
&27.7  &{\it 26.7$\pm$3.4}
\cr
STAR& 130
&265.5 &{\it 278$\pm$20}
&267.7 &{\it 280$\pm$20}
&46.7 &{\it 46.3$\pm$3.0}
&43.1 &{\it 42.7$\pm$2.8}
&28.2 &{\it 28.2$\pm$3.1}
&19.9 &{\it 20.2$\pm$2.2}	
\cr
STAR& 62.4
&221.1  &{\it 233$\pm$17}
&225.2 &{\it 237$\pm$17}
&38.3 &{\it 37.6$\pm$2.7}
&32.5 &{\it 32.4$\pm$2.3}
&29.2 &{\it 29.0$\pm$3.8}
&13.5 &{\it 13.6$\pm$1.7}
\cr
\hline
\hline
\end{tabular}
\caption{\label{tabledndy} 
Values of $dN/dy$ for identified hadrons obtained by extrapolating
measured spectra to the whole $p_T$ range. Our extrapolations are
compared 
with the extrapolations done by experimental collaborations (in italics). 
}
\end{table}

\begin{table}
\begin{tabular}{c|c|rcrcrcrcrcrc}
\hline
\hline
exp.&$\sqrt{s}$~[GeV]&$\pi^+$&&$\pi^-$&&$K^+$&&$K^-$&&$p$&&$\bar p$\cr
\hline
ALICE & 2760
&522 &{\it 517$\pm$19}
&525 &{\it 520$\pm$18}
&878 &{\it 876$\pm$26}
&867 &{\it 867$\pm$27}
&1357 &{\it 1333$\pm$33}
&1356 &{\it 1353$\pm$34}
\cr
PHENIX& 200
&438 &{\it 451$\pm$33}
&447 &{\it 455$\pm$32}
&681 &{\it 670$\pm$78}
&697 &{\it 677$\pm$68}
&1021  &{\it 949$\pm$85}
&1051 &{\it 959$\pm$84}
\cr
STAR& 200
&443 &{\it 427$\pm$22}
&437 &{\it 422$\pm$22}
&720 &{\it 720$\pm$74}
&720 &{\it 719$\pm$74}
&1102 &{\it 1104$\pm$110}
&1102 &{\it 1103$\pm$114}
\cr
STAR& 130
&414 &{\it 404$\pm$13}
&415 &{\it 404$\pm$13}
&668 &{\it 666$\pm$30}
&668 &{\it 667$\pm$30}
&1002 &{\it 1003$\pm$87}
&1002 &{\it 1002$\pm$87}
\cr
STAR& 62.4
&410 &{\it 406$\pm$11}
&407 &{\it 403$\pm$11}
&646 &{\it 646$\pm$29}
&646 &{\it 645$\pm$29}
&960 &{\it 956$\pm$75}
&960 &{\it 959$\pm$60}
\cr
\hline
\hline
\end{tabular}
\caption{\label{tablemeanpt} 
Values of $\langle p_t\rangle$ (in MeV/$c$) for identified hadrons obtained by
extrapolating measured spectra to the whole $p_T$ range. Our extrapolations are
compared 
with the extrapolations done by experimental collaborations (in italics). 
}
\end{table}
\twocolumngrid

\begin{table}
\begin{tabular}{c|c|rrrrrr}
\hline
\hline
exp.&$\sqrt{s}$~[GeV]&$\pi^+$&$\pi^-$&$K^+$&$K^-$&$p$&$\bar p$\cr
\hline
ALICE & 2760
&553
&555
&1043
&1034
&1702
&1702
\cr
PHENIX& 200
&472
&481
&878
&889
&1435
&1455
\cr
STAR& 200
&475
&470
&906
&906
&1496
&1496
\cr
STAR& 130
&448
&449
&861
&861
&1416
&1416
\cr
STAR& 62.4
&444
&441
&843
&843
&1384
&1384
\cr
\hline
\hline
\end{tabular}
\caption{\label{tablemeanmt} 
Values of $\langle m_t\rangle$ (in MeV) for identified hadrons obtained by
extrapolating measured spectra to the whole $p_T$ range. 
}
\end{table}


\end{document}